\documentclass[twocolumn,showpacs,prl]{revtex4}% Physical Review Letters

\usepackage{graphicx}% Include figure files
\usepackage{dcolumn}% Align table columns on decimal point
\usepackage{bm}% bold math

\begin{document}

\title{Superconductivity from repulsion:\\ Ginzburg~--~Landau phenomenology of cuprates}

\author{V.~I.~Belyavsky and Yu.~V.~Kopaev}

\address {P.~N.~Lebedev Physical Institute, Russian Academy of
Sciences, Moscow, 119991, Russia}

\begin{abstract}

We develop Ginzburg-Landau approach to the problem of
superconducting pairing with large momentum under screened Coulomb
repulsion (${\eta}^{}_K$~-~pairing). Two-component order parameter
arising in this scheme can be associated with charge and orbital
current degrees of freedom of the relative motion of
${\eta}^{}_K$~-~pair corresponding to superconducting and orbital
antiferromagnetic ordered states, respectively. All basic features
of the phase diagram of cuprate superconductors result directly
from the ${\eta}^{}_K$~-~pairing concept.

\vspace{0.2cm}


\end{abstract}


\maketitle

%\section{Introduction}

\noindent {\bf{1. Introduction}}

\vspace{0.2cm}

Since the discovery of superconductivity (H.~Kamerling Onnes,
1911) a search of materials with high superconducting (SC) phase
transition temperature $T^{}_c$ has been at the focus of solid
state physics. When the microscopic BCS theory of
superconductivity was developed (J.~Bardeen, L.~N.~Cooper, and
J.~R.~Schrieffer, 1957; N.~N.~Bogoliubov, 1958), it became clear
that there were some restrictions to SC transition temperature
$T^{}_c$ (M.~L.~Cohen and P.~W.~Anderson, 1972) following from the
BCS theory. According to BCS theory, the SC state arises when, due
to phonon-mediated attraction (H.~Fr\"ohlich, 1950; J.~Bardeen,
1950), electrons pair up into Cooper pairs (L.~N.~Cooper, 1956)
which, at $T<T^{}_c$, can form a coherent state (SC condensate)
insensitive to crystal imperfections. One of the restrictions to
$T^{}_c$ following from the phonon scenario of SC pairing is
connected with the fact that Debye energy ${\hbar}{\omega}^{}_D$
is much less than Fermi energy $E^{}_F$.

It was proposed (W.~A.~Little, 1964; V.~L.~Ginzburg, 1964) that,
in a low-dimensional electron system, $T^{}_c$ might be
considerably higher if pairing attraction were due to non-phonon
origin. Such an exciton-mediated attraction, just as another ways
of an increase of $T^{}_c$, were theoretically studied in detail
by Ginsburg's group in Moscow \cite{book}. A great many of the
results obtained by this group, in particular, a possibility of
superconductivity in three-dimensional insulating crystals (a
coexistence of SC and insulating states in one-dimensional
crystals was previously considered in \cite{BGD}), turned out to
be especially actual after the discovery of high temperature
superconductivity (HTSC) of doped cuprate oxides (J.~G.~Bednorz
and K.~A.~M\"uller, 1986).

Taking account of Coulomb repulsion in the framework of the phonon
mechanism of the BCS theory (V.~V.~Tolmachev, 1958) leads to a
conclusion that Cooper pairing occurs if effective coupling
constant $V$ of phonon-mediated pairing attraction exceeds
logarithmically suppressed average value of Coulomb repulsion
energy $U$:
\begin{equation}\label{1}
V>{\frac{U}{1+gU{\ln{(E^{}_F/{\hbar}{\omega}^{}_D)}}}}.
\end{equation}
Here, $g$ is density of states on the Fermi level.

At the same time, there was investigated SC pairing under
repulsive interaction itself. Such a pairing is possible if the
expansion of repulsive pairing potential into the series over
spherical harmonics contains at least one term with angular
momentum $l\neq 0$ corresponding to negative scattering length
(L.~D.~Landau, 1958). In a degenerate electron gas, this condition
is satisfied (W.~Kohn and J.~M.~Luttinger, 1964) due to Kohn
singularity of screened Coulomb repulsion. In two-band
superconductor, repulsion-induced SC pairing can be effective as
well (V.~A.~Moskalenko, 1959; H.~Suhl, B.~T.~Matthias, and
L.~R.~Walker, 1959) when there is the Suhl inequality (inverse
Cauchy-Bunyakovsky inequality),
\begin{equation}\label{2}
U^{}_{12}U^{}_{21}-U^{}_{11}U^{}_{22}>0,
\end{equation}
between interband ($U^{}_{12}=U^{}_{21}$) and intraband
($U^{}_{11}$, $U^{}_{22}$) matrix elements of the pairing
interaction.

The discovery of HTSC in cuprates has resulted in not only a
considerable increase of $T^{}_c$ but put some fundamental
problems in the theory of superconductivity. Anderson
\cite{Anderson} has pointed out general statements of physics of
cuprates: 1) these compounds are quasi-two-dimensional (2D)
systems with considerably strong electron correlations in
${\text{CuO}}^{}_2$ layers; 2) superconductivity arises under
doping of parent spin antiferromagnetic (AF) insulator. To explain
unconventional properties of cuprates, along with phonon-mediated
attraction, AF magnon-mediated SC repulsion (N.~F.~Berk and
J.~R.~Schrieffer, 1966) and direct Coulomb repulsion (most
commonly, in the framework of Hubbard model or related models
\cite{Dagotto}) are considered as underlying pairing interactions.

A choice of the microscopic mechanism of SC pairing determines the
symmetry and internal structure of SC order parameter which, in
the cuprates, may be essentially other than simple one-component
complex order parameter following from phonon scenario. It is
quite natural to consider screened Coulomb repulsion as a
reasonable underlying pairing interaction that determines both
insulating and SC states of the cuprates. This mechanism results
in the order parameter which has no less than two complex
components \cite{BKS} that, under a description \cite{BK} in the
framework of Ginzburg-Landau phenomenology (developed previously
to the microscopic BCS theory: V.~L.~Ginzburg and L.~D.~Landau,
1950), should be determined by an equation system that can admit
of more than BCS-like solution only.

In this paper, using a generalized Ginzburg-Landau approach, we
present the concept of Coulomb SC pairing with large momentum
(${\eta}^{}_K$~-~pairing, by analogy with Yang's
${\eta}$~-~pairing \cite{Yang}) and some of its applications to
physics of cuprates \cite{BK_06}.

%\section{Repulsion-induced superconducting pairing with large
%momentum}

\vspace{0.2cm}

\noindent {\bf{2. Repulsion-induced superconducting pairing with
large momentum}}

\vspace{0.2cm}

\begin{figure}
\includegraphics[]{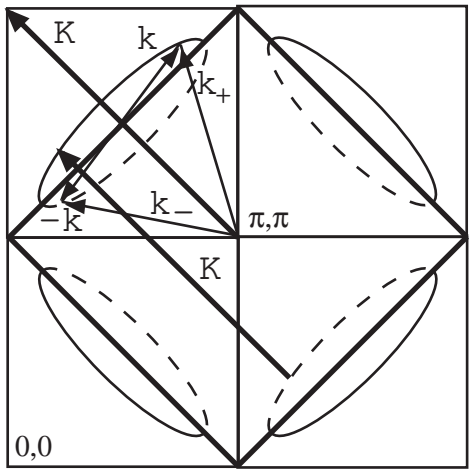}% Here is how to import EPS art
\caption[*]{Nesting and mirror nesting features of the FC typical
of cuprates (schematically). Main and shadow bands of hole pockets
are presented as solid and dashed lines, respectively. Boldface
line shows the boundary of the magnetic Brillouin zone. It is
shown how two particles with the momenta
${\bm{k}}^{}_{\pm}={\bm{K}}/2 \pm {\bm{k}}$ form a pair with total
momentum ${\bm{K}}$.}\label{Fig1.eps}
\end{figure}

Angle-resolved photoemission spectroscopy (ARPES) study
\cite{Shen,{Damascelli}} results in a conclusion that, in
cuprates, Fermi contour (FC) is situated in an extended vicinity
of saddle point of one-band 2D electron dispersion
${\varepsilon}({\bm{p}})$. At doping $x$, a {\it{large}} FC bounds
electron-filled region of the momentum space which is proportional
to $(1-x)$ and has a typical form of a {\it{square with rounded
corners}}. Spin AF insulating gap of the corresponding parent
compound survives under relatively low doping due to short-range
AF order decreasing with an increase of $x$. Therefore a part of
hole filling proportional to $x$ occupies {\it{small hole
pockets}} in the lower electron subband that are centered at the
points (${\pm{\pi}/2},{\pm{\pi}/2}$) on the boundary of the
magnetic Brillouin zone of the parent compound. Relatively high
spectral weight corresponds to the part of a pocket ({\it{main
band}}) situated inside the first magnetic zone (this part belongs
to the {\it{large}} FC) whereas the spectral weight corresponding
to the other part ({\it{shadow band}}; A. Kampf and J.R.
Schrieffer, 1990) turns out to be considerably suppressed and
decreases with $x$. Low energy quasiparticle excitations can arise
only due to hole pockets resulting, after the transition into SC
state, in low phase stiffness, ${\rho}^{}_{sf}\sim x$.

The FC in the form of hole pockets (Fig.1) possesses two special
features. Firstly, it manifests perfect {\it{nesting}} feature,
\begin{equation}\label{3}
{\varepsilon}({\bm{K}}+{\bm{p}})+{\varepsilon}({\bm{p}}) =2{\mu},
\end{equation}
($\mu$ is the chemical potential) at the momentum ${\bm{K}}
=({\pi},{\pi})$ connecting the maim and shadow parts of two
antipodal pockets as it is shown in Fig.1. Nesting of the FC can
promote a pairing in a particle-hole channel resulting in an
insulating gap on the FC. Secondly, such a FC manifests perfect
{\it{mirror nesting}} feature \cite{BKS},
\begin{equation}\label{4}
{\varepsilon}({\bm{K}}/2+{\bm{k}})={\varepsilon}({\bm{K}}/2-{\bm{k}}),
\end{equation}
at the same momentum promoting pairing in a particle-particle
channel that can result in a rise of SC condensate. In this case,
the momenta ${\bm{K}}/2 \pm {\bm{k}}$ of the particles composing a
pair with total momentum ${\bm{K}}$ (${\eta}^{}_K$~-~pair) belong
to the main and shadow parts of the same pocket (${\bm{k}}$ is a
momentum of the relative motion of ${\eta}^{}_K$~-~pair).

In the case of Cooper pairing with ${\bm{K}}=0$, the mirror
nesting condition (\ref{4}) is satisfied trivially on the whole of
any FC due to general feature of electron dispersion,
${\varepsilon}(-{\bm{p}})= {\varepsilon}({\bm{p}})$. If
${\bm{K}}\neq 0$, the momenta ${\bm{k}}^{}_{\pm} ={\bm{K}}/2 \pm
{\bm{k}}$ of the particles composing ${\eta}^{}_K$~-~pair belong
to a domain of kinematic constraint (P. Fulde and R.A. Ferrel,
1964; A.I. Larkin and Yu.N. Ovchinnikiov, 1964) being only a part
of the crystal Brillouin zone. In the case of pocket-like FC, such
a domain is a quarter of the Brillouin zone. Note that the
spectral weight corresponding to a shadow band of the pocket
decreases rapidly with an increase of a distance from the FC.
There are four crystal equivalent domains of kinematic constraint
${\Xi}^{}_j$ that can be indexed by a label $j=1,2,3,4$.

Under mirror nesting condition, the mean-field self-consistency
equation that determines SC energy gap parameter
${\Delta}({\bm{k}})$ can be written as
\begin{equation}\label{5}
{\Delta}({\bm{k}})=-{\frac{1}{2}}\sum_{{\bm{k}}_{}^{\prime}}
{\frac{U({\bm{k}}-{\bm{k}}_{}^{\prime}){\Delta}({\bm{k}}_{}^{\prime})}
{E({\bm{k}}_{}^{\prime})}} h({\bm{k}}_{}^{\prime};T)
\end{equation}
where the summation has to be performed over
${\bm{k}}_{}^{\prime}$ inside one of the domains of kinematic
constraint. Here, $U({\bm{k}}_{}^{}-{\bm{k}}_{}^{\prime})$ is
pairing interaction matrix element,
\begin{equation}\label{6}
E({\bm{k}}_{}^{})={\sqrt{{\xi}_{}^2({\bm{k}}_{}^{})
+|{\Delta}({\bm{k}}_{}^{})|_{}^2}}
\end{equation}
is quasiparticle energy,
\begin{equation}\label{7}
2{\xi}({\bm{k}}_{}^{})={\varepsilon} ({\bm{k}}_{+}^{})
+{\varepsilon}({\bm{k}}_{-}^{})-2{\mu}
\end{equation}
is kinetic energy of ${\eta}^{}_K$~-~pair, and
\begin{equation}\label{8}
h({\bm{k}};T)=\tanh [E({\bm{k}})/2T].
\end{equation}
One has to have in mind that the gap parameter in (\ref{5}) is
defined inside the domain ${\Xi}^{}_j$ corresponding to the total
momentum ${\bm{K}}^{}_j$ of ${\eta}^{}_K$~-~pair. Therefore, to
take it into consideration, it is convenient to write this
parameter in the explicit form, ${\Delta}^{}_j({\bm{k}})$.

Repulsive interaction can result in SC pairing if and only if the
kernel $U({\bm{k}}-{\bm{k}}_{}^{\prime})$ of the integral equation
(\ref{5}) has at least one {\it{negative}} eigenvalue \cite{BKS}.
Eigenfunctions ${\varphi}^{}_s({\bm{k}})$ and eigenvalues
${\lambda}^{}_s$ of the pairing interaction operator are defined
as the solutions to the equations
\begin{equation}\label{9}
{\varphi}^{}_s({\bm{k}})={\lambda}^{}_s
\sum_{{\bm{k}}_{}^{\prime}}U({\bm{k}}-{\bm{k}}_{}^{\prime})
{\varphi}^{}_s({\bm{k}}_{}^{\prime}).
\end{equation}
One can consider complete orthonormal system of the functions
${\varphi}^{}_s({\bm{k}})$ as generic basis convenient to solve
the SC pairing problem (\ref{5}).

A nontrivial solution ${\Delta}({\bm{k}})$ to the equation
(\ref{5}), if such a solution exists at all, must change its sign
on a {\it{nodal}} line inside the domain of kinematic constraint
\cite{BKS}. As far as the arguments ${\bm{k}}$ and
${\bm{k}}_{}^{\prime}$ of the kernel
$U({\bm{k}}-{\bm{k}}_{}^{\prime})$ are defined inside this domain,
such a kernel, in the case of screened Coulomb repulsion, has to
have one negative eigenvalue with necessity \cite{BKS}. To obtain
an approximate solution to the equation (\ref{5}), it is
convenient to reduce the true nondegenerate kernel to a degenerate
one with a finite set of eigenfunctions. The simplest appropriate
degenerate kernel \cite{BKS},
\begin{equation}\label{10}
U({\bm{k}}-{\bm{k}}_{}^{\prime})=U^{}_0
[1-({\bm{k}}-{\bm{k}}_{}^{\prime})_{}^2r^2_0/2],
\end{equation}
where $U^{}_0$ and $r^{}_0$ have meaning of effective coupling
constant and screening length, respectively, corresponds to two
even and two odd (with respect to inversion transformation,
${\bm{k}}\rightarrow -{\bm{k}}$) eigenfunctions. Due to the fact
that ${\Delta}(-{\bm{k}})={\Delta}({\bm{k}})$, the gap parameter
turns out to be a linear combination of even eigenfunctions only.
Therefore, in the case of repulsion pairing (\ref{10}), one can
conclude that there is a two-component SC order.

One can introduce the SC order parameter as a wave function of
${\eta}^{}_K$~-~pair,
\begin{equation}\label{11}
{\Psi}({\bm{R}},{\bm{k}})=\sum_j {\gamma}^{}_j
e_{}^{i{\bm{K}}^{}_j{\bm{R}}} {\Psi}^{}_j({\bm{k}}),
\end{equation}
where ${\bm{R}}$ is center-of-mass radius vector,
${\Psi}^{}_j({\bm{k}}) \sim {\Delta}^{}_j({\bm{k}})$ has meaning
of a wave function of the relative motion of ${\eta}^{}_K$~-~pair
with the total momentum ${\bm{K}}^{}_j$. The coefficients
${\gamma}^{}_j$ determine a symmetry of the order parameter that
is one of two one-dimensional irreducible representations of 2D
crystal symmetry group: $A^{}_{1g}$ (extended $s$~-~wave) or
$B^{}_{1g}$ (extended $d$~-~wave). A choice of ${\gamma}^{}_j$ is
connected with the pairing interaction (for example,
phonon-mediated or AF magnon-mediated) that mixes states of
particles in different domains of kinematic constraint.

Eigenfunctions of pairing interaction operator defined in the
region ${\Xi}$ joining all domains of kinematic constraint can be
written as
\begin{equation}\label{12}
{\varphi}^{}_s({\bm{R}},{\bm{k}}) =\sum_j {\gamma}^{}_j
e_{}^{i{\bm{K}}^{}_j{\bm{R}}} {\varphi}^{}_{js}({\bm{k}}).
\end{equation}
One can represent the order parameter as
\begin{equation}\label{13}
{\Psi}({\bm{R}},{\bm{k}})=\sum_s {\Psi}^{}_s({\bm{R}})
{\varphi}^{}_s({\bm{R}},{\bm{k}})
\end{equation}
that results in the components of the order parameter in the form
of
\begin{equation}\label{14}
{\Psi}^{}_s({\bm{R}})=\sum_{{\bm{k}}\in{\Xi}}
{\Psi}({\bm{R}},{\bm{k}}) {\varphi}^{\ast}_s({\bm{R}},{\bm{k}}).
\end{equation}

\begin{figure}
\includegraphics[]{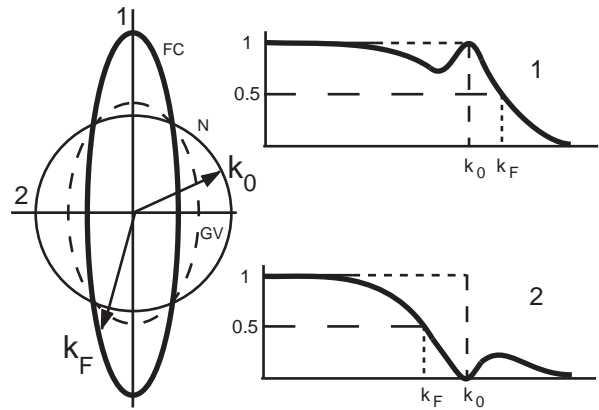}% Here is how to import EPS art
\caption[*]{Left: Three peculiar nodal curves inside the domain of
kinematic constraint: Fermi contour (FC) in the form of hole
pocket, nodal curve (N) where SC order parameter changes sign, and
the curve (GV) on which there is a minimum of quasiparticle energy
(quasiparticle group velocity becomes zero). Right: Momentum
dependence of the occupation number $v^2_{\bm{k}}$ is shown for
two directions (1,2, in accordance with the left figure) in the
momentum space.}\label{Fig2.eps}
\end{figure}

As it follows from ${\eta}^{}_K$~-~pairing concept, under mirror
nesting condition, there are three peculiar nodal curves (Fig.2)
belonging to each domain of kinematic constraint: 1) FC (or its
finite piece) on which kinetic energy of ${\eta}^{}_K$~-~pair
equals zero, $2{\xi}({\bm{k}})=0$, and quasiparticle charge
changes its sign; 2) nodal curve on which
${\Delta}^{}_j({\bm{k}})=0$ and SC order parameter changes its
sign; 3) a curve on which
${\nabla}^{}_{\bm{k}}{\Delta}^{}_j({\bm{k}})=0$ and quasiparticle
group velocity changes its sign. These three curves do not
coincide and, as a consequence, coherence factors manifest
peculiar momentum dependence as it is shown schematically in
Fig.2.

The mean-field SC transition temperature $T^{}_{sc}$,
corresponding to ${\Delta}^{}_j({\bm{k}})\rightarrow 0$, can be
derived from the self-consistency equation (\ref{5}). It should be
noted that pairing repulsion, in particular, the model potential
(\ref{10}), makes possible a rise of quasi-stationary states (QSS)
of ${\eta}^{}_K$~-~pair as solutions of corresponding Cooper
problem \cite{BKTS}. Such QSS treated as developed fluctuations of
noncoherent ${\eta}^{}_K$~-~pairs creating above $T^{}_{sc}$  can
be associated \cite{BKTS} with strong pseudogap state observed in
underdoped cuprates \cite{Xu}.

Complex two-component SC order parameter can be normalized in the
following way:
\begin{equation}\label{15}
\sum_{{\bm{k}}\in{\Xi}}|{\Psi}({\bm{R}},{\bm{k}})|_{}^2\equiv
\sum_s|{\Psi}^{}_s({\bm{R}})|_{}^2 =n^{}_{sf}({\bm{R}})/2.
\end{equation}
Here, $n^{}_{sf}({\bm{R}})\sim{\rho}^{}_{sf}$ is the superfluid
density. As it follows from (\ref{15}), the components of the
order parameter can be represented in the form
\begin{equation}\label{16}
{\Psi}^{}_1={\psi}_1e_{}^{i{\Phi}}, \quad
{\Psi}^{}_2={\psi}^{}_2e_{}^{i{\beta}}e_{}^{i{\Phi}}
\end{equation}
where ${\psi}^{}_1$ and ${\psi}^{}_2$ are the absolute values of
the components, ${\beta}$ is the relative phase of these
components and ${\Phi}$ is the phase relating to center-of-mass
motion of ${\eta}^{}_K$~-~pair. Due to the normalization condition
(\ref{15}), ${\psi}^{}_1$ and ${\psi}^{}_2$ are not independent
and, to take into account the contribution of the relative motion
of ${\eta}^{}_K$~-~pair into SC order parameter,  one needs only
two variables, for example, ${\psi}^{}_1$ and ${\beta}$. Further
we restrict ourselves to the simplest case when ${\psi}^{}_1
={\psi}^{}_2 \equiv {\psi}$.

%\section{Ginzburg-Landau approach}

\vspace{0.2cm}

\noindent {\bf{3. Ginzburg-Landau approach}}

\vspace{0.2cm}

General expression of Ginzburg-Landau functional has the form
\begin{equation}\label{17}
F=\int d^2_{}R \; [f^{}_1+f^{}_g+f^{}_m]
\end{equation}
where, in the case of complex two-component order parameter, an
expansion of free energy density in powers of the order parameter
components  should be written as
\begin{equation}\label{18}
f^{}_1=\sum_{ss'}A^{}_{ss'}{\Psi}^{\ast}_s{\Psi}^{}_{s'} +{\frac
12} \sum_{ss'tt'} B^{}_{ss'tt'}{\Psi}^{\ast}_s{\Psi}^{\ast}_{s'}
{\Psi}^{}_{t}{\Psi}^{}_{t'}.
\end{equation}
A gradient contribution into free energy density can be presented
as
\begin{equation}\label{19}
f^{}_g={\frac
{{\hbar}_{}^2}{4m}}\sum_{ss'}[{\hat{\bm{D}}}{\Psi}^{}_s]_{}^{\dag}
M^{}_{ss'} [{\hat{\bm{D}}}{\Psi}^{}_{s'}].
\end{equation}
Here, ${\hat{\bm{D}}}=-i{\nabla}-(2e/{\hbar}c){\bm{A}}$ is the
operator of covariant differentiation, ${\bm{A}}$ is vector
potential. The explicit form of the matrices $A^{}_{ss'}$,
$B^{}_{ss'tt'}$ and $M^{}_{ss'}$ following from
${\eta}^{}_K$~-~pairing scheme is obtained in \cite{BK} within
the framework of the original Gor'kov method (L.P.Gor'kov, 1959).
Hermitian $2\times 2$ matrices $A^{}_{ss'}$ and $M^{}_{ss'}$ can
be diagonalized simultaneously by a unitary transformation of the
order parameter. Here, we suppose that such a transformation is
already performed. Thus, each of the second order matrices
$A^{}_{ss'}$ and $M^{}_{ss'}$ can be characterized by its two
principal values whereas the fourth order term in $f^{}_1$ should
be described by five independent components of $B^{}_{ss'tt'}$.

Magnetic field energy density,
\begin{equation}\label{20}
f^{}_m=(z^{}_0/8{\pi})({\text{rot}}{\bm{A}})_{}^2
\end{equation}
(here, $z^{}_0$ is the separation between the neighbor cuprate
planes) is determined by an average square of magnetic field
strength ${\bm{H}}={\text{rot}}{\bm{A}}$ which can be presented as
a sum ${\bm{H}}={\bm{H}}^{}_e+{\bm{H}}^{}_i$ where ${\bm{H}}^{}_e$
and ${\bm{H}}^{}_i$ are the contributions due to external sources
and possible spontaneous magnetization of the system,
respectively. Note that an average (over a macroscopic region of
the real space) value of the internal field ${\bm{H}}^{}_i$ equals
zero (such a condition excludes any ferromagnetic order from the
consideration) whereas the average value of $H_i^2$, of course,
should not vanish.

The system of Ginzburg-Landau equations following from the
functional (\ref{17}) is derived in \cite{BK}. One can believe
that this equation system, just as similar system of equations
corresponding to a charged two-condensate Ginzburg-Landau model
\cite{Faddeev}, may admit more various solutions as compared to
the Ginzburg-Landau equations in the case of one-component order.
To show this \cite{BKSm}, let us consider here the simplest case
of spatially homogeneous system without any magnetic fields when
it is sufficient to remain only the contribution of $f^{}_1$ into
the functional (\ref{17}). One can rewrite (\ref{18}) as
\begin{equation}\label{21}
f^{}_1=a^{}_1{\psi}_{}^2+(B+2C{\cos {\beta}}+D{\cos
{\beta}}_{}^2)\; {{\psi}_{}^4}/2
\end{equation}
where $a^{}_1=A^{}_1+A^{}_2$ is the trace of the matrix
$A^{}_{ss'}$, and the coefficients $B$, $C$ and $D$ can be
expressed in terms of the independent elements of the matrix
$B^{}_{ss'tt'}$. Here $B$ and $D$ are positive by definition. For
the sake of simplicity, we assume that $C>0$ as well.

The coefficients $A^{}_1$ and $A^{}_2$ vanish at one and the same
temperature $T^{}_{sc}$ corresponding to the mean field
approximation \cite{BK}. Therefore, at $|T-T^{}_{sc}|\ll
T^{}_{sc}$, one can assume that $a^{}_1=-{\kappa}^{}_1{\tau}^{}_1$
where ${\kappa}^{}_1>0$ and ${\tau}^{}_1=(T^{}_{sc}-T)/T^{}_{sc}$.
Besides, the coefficients $B$, $C$ and $D$ can be considered as
taken at $T=T^{}_{sc}$.

However, one should take into account that, aside from a
dependence on temperature, the coefficients in (\ref{21}) are
dependent on doping level $x$. At $T<T^{}_s$, a nontrivial
solution minimizing the free energy density (\ref{21}) corresponds
to ${\psi}\neq 0$ and ${\beta}={\pi}$ when $C\geq D$. It should be
noted that the relative phase ${\beta}={\pi}$ displays the fact
that SC order parameter arising under repulsive pairing
interaction should change its sign inside the domain of kinematic
constraint. In the opposite case $C\leq D$ the relative phase of
the components of the order parameter is determined by the
equality ${\cos {\beta}}=-C/D$. The equation $C(x)=D(x)$
determines a doping level $x=x^{}_0$ corresponding to a
qualitative change of SC order.

\vspace{0.2cm}

\noindent {\bf{4. Spontaneous orbital currents}}

\vspace{0.2cm}

One can naturally define the order parameter ${\alpha}
={\pi}-{\beta}$ to distinguish two SC phases with ${\beta}<{\pi}$
(${\beta}$-phase) and ${\beta}={\pi}$ (${\pi}$-phase). We assume
that ${\pi}$-phase corresponding to ${\alpha}=0$ may exist at
$x>x^{}_0$ whereas ${\beta}$-phase with nonzero $\alpha$ occurs at
$x<x^{}_0$. In a small vicinity of the point $x=x^{}_0$,
$T=T^{}_{sc}(x^{}_0)$, the free energy density can be represented
in the form of an expansion in even powers of the order parameters
$\psi$ and $\alpha$.

A rise of the relative phase ${\beta}\neq {\pi}$ corresponding to
complex coherence factors can be associated with an internal
magnetic field-induced change in the phase of the destruction
operator of an electron with spin ${\sigma}$ on a lattice site
with radius vector ${\bm{n}}$,
\begin{equation}\label{22}
{\hat{c}}^{}_{{\bm{n}}{\sigma}} \rightarrow
{\hat{c}}^{}_{{\bm{n}}{\sigma}}\cdot
\exp{[i(e/{\hbar}c){\bm{A}}^{}_i({\bm{n}}){\bm{n}}]}.
\end{equation}
Therefore the phase of the order parameter in the real space
representation can be written as
\begin{equation}\label{23}
{\beta}({\bm{n}},{\bm{n'}})={\pi}-(e/{\hbar}c)
[{\bm{A}}({\bm{n}}){\bm{n}}+{\bm{A}}^{}_i({\bm{n'}}){\bm{n'}}].
\end{equation}
A contribution ${\Phi}=(2e/{\hbar}c){\bm{A}}^{}_i({\bm{R}})
{\bm{R}}$ into the phase (\ref{23}) may be related to
center-of-mass motion of the pair with radius vector
${\bm{R}}=({\bm{n}}+{\bm{n'}})/2$. One can assume that vector
potential ${\bm{A}}^{}_i$ of the internal magnetic field is due to
orbital currents circulating inside a unit cell as a result of the
relative motion of ${\eta}^{}_K$~-~pair. When $x\approx x^{}_0$, a
small change in the relative phase can be represented as
\begin{equation}\label{24}
{\alpha}\approx {\frac {e}{2{\hbar}c}}{\frac {\partial
A^{}_i}{\partial x^{}_k}}x^{}_ix^{}_k
\end{equation}
where summation over repeated Cartesian indices $i,k=1,2$ is
understood.

In the SC state, orbital antiferromagnetic (OAF) order
\cite{Halperin,Volkov,Ginzburg, Affleck,Varma} that can be
associated with nonzero relative phase $\alpha$ of the SC order
parameter develops as AF correlated orbital current circulations
\cite{Ivanov}. Such circulations may survive above $T^{}_c$ in the
form of either long-range \cite{CLMN} or short-range \cite{Lee}
OAF order. In the case under consideration, real magnetic field
due to orbital currents in (\ref{24}) can be associated with a
gauge field which links the charge and orbital current degrees of
freedom ($\psi$ and $\alpha$, respectively). This field is similar
to the gauge fields introduced into Ginzburg-Landau functional in
boson version of spin-charge separation scheme \cite{Muthucumar}.

In the framework of Ginzburg-Landau phenomenology, an order
parameter should be understood as averaged over the relative
motion of ${\eta}^{}_K$-pair. Therefore, taking into account a
checkerboard distribution of orbital currents, mean square value
of OAF order parameter can be estimated as
\begin{equation}\label{25}
{\alpha}_{}^2\simeq {\left ({\frac {ea}{4{\hbar}c}} \right )}_{}^2
{A_{i}^2}
\end{equation}
where $a$ is a separation between neighbor copper atoms in
cuprate plane.

Spontaneous orbital currents in the SC state lead to a positive
contribution of internal magnetic field energy into
Ginzburg-Landau functional. In spatially homogeneous system
without an {\it{external}} magnetic field, a rise of the relative
phase change $\alpha$ introduces another contribution into the
free energy resulting from the gradient term of the
Ginzburg-Landau functional \cite{BK}. One can represent this
contribution, that links two competing orders $\psi$ and $\alpha$,
as $F^{}_{12}=b^{}_{12}{\psi}_{}^2{\alpha}_{}^2$. Here we consider
a positive constant $b^{}_{12}(x)$ as a phenomenological
parameter.

These essentially positive contributions themselves exclude a
possibility of a rise of a thermal stable state with ${\alpha}\neq
0$. Therefore, one has to realize a complete study of a
competition between the SC and insulating OAF pairing channels.

\begin{figure}
\includegraphics[]{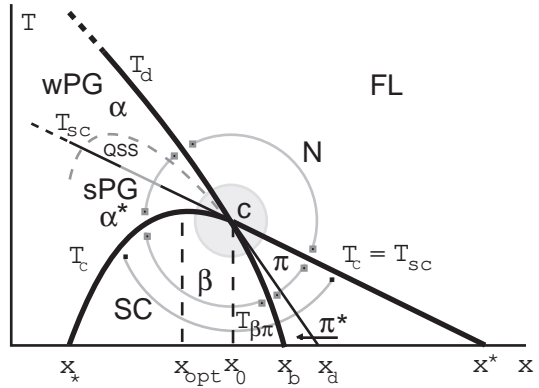}% Here is how to import EPS art
\caption[*]{The region of the phase diagram relating to a
competition of SC and OAF states in the vicinity of the
tetracritical point $c$. Bold lines show second order phase
transition curves. Weak pseudogap (wPG) and strong pseudogap (sPG)
correspond to long-range insulating order (${\alpha}$-phase) with
developed SC fluctuations (in the form of QSS of
${\eta}^{}_K$~-~pairs) in the sPG state (excited state
${\alpha}_{}^{\ast}$ of the ${\alpha}$-phase). Dashed line
restricts the region of the existence of QSS and can be related to
a crossover between the sPG and wPG states. The SC dome consists
of two phases with long-range SC order: ${\beta}$-phase
corresponding to a coexistence of SC and OAF orders and
conventional ${\pi}$-phase including the region of developed
fluctuations of orbital circular currents (excited state
${\pi}_{}^{\ast}$ of the ${\pi}$-phase).}\label{Fig1.eps}
\end{figure}

Thus, Landau free energy density of the OAF insulating state,
\begin{equation}\label{26}
f^{}_2=a^{}_2\; {\alpha}_{}^2+ b^{}_2\; {{\alpha}_{}^4}/2,
\end{equation}
should be added into the Ginzburg-Landau functional. Here,
$b^{}_2$ is a positive function of doping and the coefficient
$a^{}_2$ vanishing at the mean-field OAF transition temperature
$T^{}_d(x)$, under the condition that $|{\tau}^{}_2| \ll 1$, can
be written as $a^{}_2=- {\kappa}_{2}^{}{\tau}^{}_2$ where
${\kappa}_{2}^{}>0$, ${\tau}^{}_2 =(T^{}_d-T)/T^{}_d$. The
magnetic field energy of circulating currents, being also
proportional to ${\alpha}_{}^2$, is presupposed to be included
into the term $a^{}_2{\alpha}_{}^2$. Without OAF ordering, the
thermal stable SC ${\beta}$~--~phase turns out to be impossible
and, in such a case, the condition $C(x)=D(x)$ should be
understood as the equation which determines the lower boundary
$x^{}_{\ast}$ of the SC dome with the only one possible
${\pi}$~--~phase.

The free energy density of a homogeneous state, up to the terms of
the fourth order, can be written as
\begin{equation}\label{27}
f=a^{}_1\; {\psi}_{}^2 +a^{}_2\; {\alpha}_{}^2 +{\frac 12}\;
b^{}_1\; {\psi}_{}^4 +b^{}_{12}\; {\psi}_{}^2\; {\alpha}_{}^2
+{\frac 12}\; b^{}_2\; {\alpha}_{}^4.
\end{equation}
The conditions that $a^{}_1(T,x)=0$, $a^{}_2(T,x)=0$ determine the
mean-field temperatures $T^{}_{sc}(x)$ and $T^{}_{d}(x)$,
respectively. Doping suppresses both OAF and SC orders. At low
doping, OAF order may dominate SC one. Therefore, one can assume
that there is an intersection of the functions $T^{}_d(x)$ and
$T^{}_s(x)$ at a point corresponding to a doping level $x=x^{}_0$
as it is shown in Fig.3. It should be emphasized that the
expression (\ref{27}) represents the free energy inside relatively
small region of the phase diagram where both $|{\tau}^{}_1|\ll 1$
and $|{\tau}^{}_2|\ll 1$.

\vspace{0.2cm}

\noindent {\bf{5. Phase diagram}}

\vspace{0.2cm}

Free energy density expansion in powers of $\psi$ and $\alpha$,
Eq.(\ref{27}), is justified in a relatively small vicinity of the
intersection point $T^{}_{tc},x^{}_0$ of the functions
$T=T^{}_d(x)$ and $T=T^{}_{sc}(x)$. At this point
({\it{tetracritical point}} $c$ in Fig.3), four curves of second
order phase transitions come together. In Fig.3, these curves
(which, under the approximations we use, may terminate only on the
coordinate axes of the $T-x$ phase diagram) are prolonged outside
of the tetracritical point vicinity, however, such an extension
which does not directly follow from (\ref{27}) should be
considered as a very conditional one.

\begin{figure}
\includegraphics[]{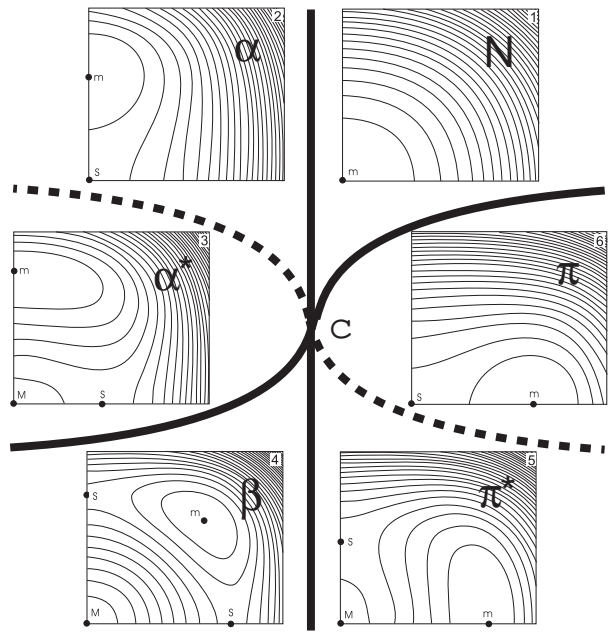}% Here is how to import EPS art
\caption[*]{Sketch of isolines of the free energy density
Eq.(\ref{27}) in different regions of the phase diagram shown in
Fig.3. Different phases are separated by bold lines. Dashed lines
bound the regions of developed fluctuations of the order
parameter.}\label{Fig4.eps}
\end{figure}

Sector $N$ in Fig.3 is related to normal Fermi liquid state
($N$-phase corresponding to high values of both temperature and
doping). Lowering of temperature or doping may result in second
order transition from $N$-phase either into insulating
$\alpha$-phase at $T=T^{}_d$ when $T>T^{}_{tc}$, $x<x^{}_0$ or
into SC $\pi$-phase at $T=T^{}_{sc}$ when $T<T^{}_{tc}$,
$x>x^{}_0$. A subsequent lowering of temperature leads to second
order transition from $\alpha$-phase into SC $\beta$-phase at
$T^{}_c(x)<T^{}_{sc}(x)$. The curve $T=T^{}_{{\beta}{\pi}}(x)$ of
second order transition between $\beta$ and $\pi$ phases inside
the SC state is situated below $T^{}_d(x)$, starts from the
tetracritical point and terminates at a point $T=0, x=x^{}_b$ on
the $T$-axis. Thus, the point $T=0$, $x=x^{}_b$ has meaning of a
quantum critical point.

At $x<x^{}_0$, the mean-field temperature $T^{}_{sc}$ is not a
phase transition temperature, similar to the mean-field
temperature $T^{}_d$ which cannot be a phase transition
temperature at $x>x^{}_0$. However, in the regions
$T^{}_c<T<T^{}_{sc}$, $x<x^{}_0$ of the insulating $\alpha$-phase
and $T^{}_{{\beta}{\pi}}<T<T^{}_d$, $x>x^{}_0$ of SC $\pi$-phase,
there arise peculiar states corresponding to developed
fluctuations of the absolute value $\psi$ and relative phase
$\alpha$ of the order parameter, respectively.

In Fig.4, we present characteristic structure of isolines of the
free energy density $f({\psi}, {\alpha})$, Eq.(\ref{27}), in
different regions of the phase diagram shown in Fig.3. In the case
of $N$-phase, there is the only singular point, namely, a
{\it{minimum}} at ${\psi} ={\alpha} =0$. In insulating
${\alpha}$-phase, there is a {\it{minimum}} at ${\psi}=0$,
${\alpha}\neq 0$ whereas the origin of coordinates ${\psi}
={\alpha} =0$ corresponds to a {\it{saddle point}}. Similarly, in
SC ${\pi}$-phase, there are a {\it{minimum}} at ${\psi}\neq 0$,
${\alpha}= 0$ and a {\it{saddle point}} at ${\psi} ={\alpha} =0$.
A transition from ${\alpha}$-phase or ${\pi}$-phase into
${\beta}$-phase leads to a qualitative change in a distribution of
singular points. First of all, the saddle point existing in both
${\alpha}$ and ${\pi}$ phases turns out to be displaced from the
origin of coordinates (in this case, the point ${\psi} ={\alpha}
=0$ becomes a {\it{local maximum}}) along the $\psi$-axis in
$\alpha$-phase (at $T^{}_c<T<T^{}_{sc}$) or along the
$\alpha$-axis in $\pi$-phase (at
$T^{}_{{\beta}{\pi}}<T<T^{}_{d}$), respectively. Then, after a
transition into $\beta$-phase at $T^{}_c$ or
$T^{}_{{\beta}{\pi}}$, there emerge a {\it{minimum}} at nonzero
both $\psi$ and $\alpha$, {\it{two saddle points}} at ${\psi}\neq
0$, ${\alpha}= 0$ and ${\psi}= 0$, ${\alpha}\neq 0$, and also a
{\it{local maximum}} at the origin of coordinates.

In the ${\alpha}$-phase at $T^{}_c<T<T^{}_{sc}$, the energies of
the minimum and saddle point turn out to be close to each other.
Therefore this region of the phase diagram corresponds to
developed fluctuations of the absolute value $\psi$ of the order
parameter. Such fluctuations can be described as creation and
annihilation of QSS of noncoherent ${\eta}^{}_K$~-~pairs. At
$x<x^{}_0$, the temperature $T^{}_{sc}$ should be related to a
crossover between the states of long-range insulating OAF order in
the ${\alpha}$-phase and developed SC fluctuations inside this
phase (an excited state ${\alpha}_{}^{\ast}$ of the
${\alpha}$-phase). In this connection, it should be noted that the
crossover temperature as an upper boundary of strong pseudogap
state may be found considerably higher than $T^{}_{sc}$ because
${\eta}^{}_K$~-~pairing admits QSS \cite{BKTS} which may arise
above the mean-field temperature $T^{}_{sc}$ (Fig.3).

The transition from ${\beta}$-phase into ${\pi}$-phase inside the
SC state also occurs through a region
($T^{}_{{\beta}{\pi}}<T<T^{}_d$, $x^{}_0<x<x^{}_d$) of developed
fluctuations as it is seen from Fig.3. In this case, the relative
phase $\alpha$ turns out to be a fluctuating parameter and
$T^{}_d(x)$ is a crossover temperature. One can imagine that QSS
in the form of noncorrelated circular orbital currents may exist
inside the region of such fluctuations (${\pi}_{}^{\ast}$, in
Fig.3). At last, SC ${\beta}$-phase corresponds to a coexistence
of superconductivity and orbital antiferromagnetism.

Phase transition in such quasi-two-dimensional systems as cuprates
has to be similar to Berezinski~-~Kosterlitz~-~Thouless (BKT)
transition (V.L.~Berezinski, 1971; J.M.~Kosterlitz and
D.J.~Thouless, 1973) which can be described by the thermal
unbinding of vortex-antivortex pairs. In contrast to vortex
lifetime in nonsuperconducting state of normal metal, BKT
transition temperature $T^{}_{BKT}$ is weakly sensitive to vortex
core energy. Therefore, in the case of transition from the SC
state into normal metal state, unbinding vortices survive in a
very narrow temperature interval above $T^{}_{BKT}$. We believe
that such a case corresponds to the transition between $N$ and
${\pi}$-pases.

One can think that, in underdoped and slightly overdoped cuprates,
vortex core may be related not to a normal metal but to an
insulating state competing with the SC state. It leads to highly
low vortex core energy \cite{Lee} so that vortex excitations (as
QSS of noncoherent ${\eta}^{}_K$~-~pairs) may exist in rather wide
temperature range which can be associated with the strong
pseudogap state. It should be noted here that magnetic
field-induced destruction of superconductivity in cuprates results
in exactly insulating state \cite{Morgan}.

In the framework of the BKT scenario, loss of SC phase coherence
is due to a rise of Abrikosov vortices as topological defects of
the phase $\Phi$ relating to the center-of-mass motion of
${\eta}^{}_K$~-~pair. One can think that, in this scenario, the
transition ${\beta}\rightarrow {\pi}$ is also accompanied by
thermal excitation of vortices. As far as center-of-mass phase
$\Phi ={\text{const}}$ in the SC state, such vortices should be
manifested in the relative motion of ${\eta}^{}_K$~-~pair arising
due to unbinding of bounded orbital current circulations of
opposite sign. It should be emphasized especially that developed
fluctuations arising in the mean-field scheme considered here
emerge due to a competition of two ordered states and peculiar
features of ${\eta}^{}_K$~-~pairing under repulsive interaction.

\vspace{0.2cm}

\noindent {\bf{5. Conclusion}}

The ${\eta}^{}_K$~-~pairing concept presented here in the
framework of a generalized Ginzburg-Landau approach seems quite
adequate for the explanation of the phase diagram typical of SC
cuprate compounds. Such a concept leads naturally to the relative
phase of the two-component SC order parameter which can be
associated with an insulating OAF ({\it{hidden}} \cite{CLMN})
order. Thus, a competition of the SC and OAF ordered states turns
out to be a generic input principle of the macroscopic approach to
the problem of superconductivity of cuprates.

All principal features of the phase diagram of the cuprates can be
qualitatively described in the framework of the
${\eta}^{}_K$~-~pairing concept: weak and strong pseudogaps, the
form of the SC dome, the regions of developed fluctuations of the
SC and OAF components of the order parameter. This concept
predicts a possibility of the second order phase transition inside
the SC dome: one of the two SC phases corresponds to a coexistence
of superconductivity and AF ordered circulations of orbital
currents whereas the other one is similar to conventional SC
phase. An enhanced diamagnetic response and giant Nernst effect
\cite{Corson} observed in underdoped cuprates in a broad
temperature region above $T^{}_c$ can also be entered into the
${\eta}^{}_K$~-~pairing scheme \cite{BKSm}.

Screened Coulomb repulsion as an underlying pairing interaction of
the ${\eta}^{}_K$~-~pairing scheme allows to explain reasonably
both the SC and insulating behavior of the cuprates. In
particular, this scheme can qualitatively explain the observed
evolution of the form of the FC with doping \cite{BKK}. Main and
shadow parts of small hole pockets manifest perfect nesting and
mirror nesting features of the FC at the same momentum
${\bm{K}}=({\pi},{\pi})$. These nesting features provide a
possibility of a rise of both insulating and SC states at
arbitrarily value of a coupling constant (asymptotic exact
solution to the mean-field pairing problem).

The Drude-like behavior of the optical conductivity \cite{Basov}
in the SC state of the cuprates up to extremely low temperature
can be treated, in the framework of the ${\eta}^{}_K$~-~pairing
concept, as the fact that a considerable part of carriers does not
belong to the SC condensate. In overdoped and nearly optimal doped
cuprates, only a piece of the FC exhibits the mirror nesting
feature. Therefore, the superfluid density is proportional to the
length of this piece whereas the normal density remains finite up
to the zero-temperature limit. In underdoped cuprates, the FC has
the form of small hole pockets with perfect mirror nesting on the
whole of each pocket. However, the considerable difference between
the values of the spectral weight on the two opposite sides of the
pocket (main and shadow bands) results in the fact that the SC
pairing turns out to be governed by the shadow band with low
spectral weight: only a small part of carriers in the main band
can find the partners in the shadow band. Low-temperature heat
capacity, $C^{}_V\sim {\gamma}T$ \cite{Loram}, is also determined
by such off-condensate carriers.

Nontrivial momentum dependence of the coherence factors resulting
from the ${\eta}^{}_K$~-~pairing scheme allows to explain observed
tunnel conductance asymmetry, {\it{peak and dip-hump}} structure
of both tunnel and photoemission spectra \cite{Kopaev_Sofronov},
and typical of cuprates features \cite{Deutscher} of Andreev
reflection \cite{Kopaev_Sofronov1}.

The ${\eta}^{}_K$~-~pairing scheme, extended to account for
phonon-mediated pairing, allows to explain \cite{BKNT}
qualitatively observed behavior of the isotope effect in different
cuprates including both absence of this effect in some compounds
\cite{Tallon} and the so-called negative isotope effect
\cite{Frank_Lawrie}.

A competition of screened Coulomb pairing repulsion and
phonon-mediated attraction or AF magnon-mediated repulsion
determines \cite{BKNT} observed unconventional symmetry of the SC
order parameter in cuprates \cite{Zhao,Brandow}.

\vspace{0.2cm}

\vspace{0.1cm}

This work was supported in part by the Russian Foundation for
Basic Research (project nos. 05-02-17077, 06-02-17186).

\end{document}